\documentclass[aps,prl,twocolumn,superscriptaddress]{revtex4-1}

\usepackage{graphicx,latexsym}
\usepackage{amsmath,amssymb,amsfonts}
\usepackage{bm}
\usepackage{braket}

\begin{document}


\title{Exact Liouvillian Spectrum of a One-Dimensional Dissipative Hubbard Model}


\author{Masaya Nakagawa}
\email{nakagawa@cat.phys.s.u-tokyo.ac.jp}
\affiliation{Department of Physics, University of Tokyo, 7-3-1 Hongo, Bunkyo-ku, Tokyo 113-0033, Japan}
\author{Norio Kawakami}
\affiliation{Department of Physics, Kyoto University, Kyoto 606-8502, Japan}
\author{Masahito Ueda}
\affiliation{Department of Physics, University of Tokyo, 7-3-1 Hongo, Bunkyo-ku, Tokyo 113-0033, Japan}
\affiliation{RIKEN Center for Emergent Matter Science (CEMS), Wako, Saitama 351-0198, Japan}
\affiliation{Institute for Physics of Intelligence, University of Tokyo, 7-3-1 Hongo, Bunkyo-ku, Tokyo 113-0033, Japan}


\date{\today}

\begin{abstract}
A one-dimensional dissipative Hubbard model with two-body loss is shown to be exactly solvable. We obtain an exact eigenspectrum of a Liouvillian superoperator by employing a non-Hermitian extension of the Bethe-ansatz method. We find steady states, the Liouvillian gap, and an exceptional point that is accompanied by the divergence of the correlation length. A dissipative version of spin-charge separation induced by the quantum Zeno effect is also demonstrated. Our result presents a new class of exactly solvable Liouvillians of open quantum many-body systems, which can be tested with ultracold atoms subject to inelastic collisions.
\end{abstract}


\maketitle


In quantum physics, no realistic system can avoid the coupling to an environment. 
The problem of decoherence and dissipation due to an environment is crucial even for small quantum systems. Furthermore, recent remarkable progress in quantum simulations with a large number of atoms, molecules, and ions has raised a fundamental and practical problem of understanding open quantum many-body systems, where interparticle correlations are essential \cite{Mueller12, Daley14, Weimer10, Barreiro11}. 
Within the Markovian approximation, the nonunitary dynamics of an open quantum system is generated by a Liouvillian superoperator acting on the density matrix of the system \cite{Lindblad76, GKS76, BreuerPetruccione}. 
While interesting solvable examples have been found \cite{Prosen08, Prosen11, Karevski13, Prosen14, Medvedyeva16, Banchi17, Rowlands18, Ribeiro19, Shibata19_1, Shibata19_2, Ziolkowska19}, the diagonalization of a Liouvillian of a quantum many-body system is more challenging than that of a Hamiltonian. 
Extending the class of exactly solvable models to the realm of dissipative systems and discovering prototypical solvable models that can be realized experimentally should promote the deepening of our understanding of strongly correlated open quantum systems. 

The Hubbard Hamiltonian provides a quintessential model in quantum many-body physics, where the interplay between quantum-mechanical hopping and interactions plays a key role. In particular, equilibrium properties of the one-dimensional case are well understood with the help of the exact solutions \cite{LiebWu68, LiebWu03, 1dHubbard_book}. 
The Hubbard model has been experimentally realized with ultracold fermionic atoms in optical lattices \cite{Esslinger10}, and the high controllability in such systems has recently invigorated the investigation of the effect of dissipation due to particle losses \cite{Sponselee18}. 
In this Letter, we show that the one-dimensional Hubbard model subject to two-body particle losses is exactly solvable. 
On the basis of the exact solution, we obtain an eigenspectrum of the Liouvillian, and elucidate how dissipation fundamentally alters the physics of the Hubbard model. 
Our main findings are threefold. 
First, we obtain the exact steady states and long-lived eigenmodes that govern the relaxation dynamics after a sufficiently long time. 
Second, we show that excitations above the Hubbard gap are significantly altered by dissipation and find that the model shows novel critical behavior near an exceptional point \cite{Heiss12} that originates from the nondiagonalizability of the Liouvillian. 
Third, we demonstrate that spin-charge separation, which is a salient feature of one-dimensional systems \cite{Giamarchi_book}, is extended to dissipative systems by exploiting the fact that the strong correlation is induced by dissipation even in the absence of an interaction. 
Our result shows that a number of exactly solvable Liouvillians can be constructed from quantum integrable models subject to loss.


\textit{Setup}.--\ 
We consider an open quantum many-body system described by a quantum master equation in the Gorini-Kossakowski-Sudarshan-Lindblad form \cite{Lindblad76, GKS76, BreuerPetruccione}
\begin{equation}
\frac{d\rho}{d\tau}=-i[H,\rho]+\sum_{j=1}^L(L_j\rho L_j^\dag-\frac{1}{2}\{ L_j^\dag L_j, \rho\})\equiv\mathcal{L}\rho,
\label{eq_master}
\end{equation}
where $\rho(\tau)$ is the density matrix of a system at time $\tau$. 
The system Hamiltonian $H$ is given by the Hubbard model on an $L$-site chain
\begin{equation}
H=-t\sum_{j=1}^L\sum_{\sigma=\uparrow,\downarrow}(c_{j,\sigma}^\dag c_{j+1,\sigma}+\mathrm{H.c.})+U\sum_{j=1}^L n_{j,\uparrow}n_{j,\downarrow},
\label{eq_Hubbard}
\end{equation}
where $c_{j,\sigma}$ is the annihilation operator of a spin-$\sigma$ fermion at site $j$, and $n_{j,\sigma}\equiv c_{j,\sigma}^\dag c_{j,\sigma}$. 
The Lindblad operator $L_j=\sqrt{2\gamma}c_{j,\downarrow}c_{j,\uparrow}$ describes a two-body loss at site $j$ with rate $\gamma>0$, which is caused by on-site inelastic collisions between fermions as observed in cold-atom experiments \cite{Syassen08, Tomita17, Tomita18, Sponselee18}. 
The formal solution of the quantum master equation can be written down in terms of the eigensystem of the Liouvillian superoperator $\mathcal{L}$ defined in Eq.~\eqref{eq_master}. 
In this Letter, we aim at diagonalizing the Liouvillian and obtain exact results for the effect of dissipation on correlated many-body systems.


\textit{Diagonalization of the Liouvillian}.--\ 
The one-dimensional Hubbard model, Eq.~\eqref{eq_Hubbard}, is known to be solvable with the Bethe ansatz \cite{LiebWu68, LiebWu03, 1dHubbard_book}. 
Here, we generalize the solvability of the Hubbard Hamiltonian to that of the Liouvillian on the basis of the existence of a conserved quantity in the Hamiltonian \cite{Torres14}. 
We first decompose the Liouvillian into two parts as $\mathcal{L}=\mathcal{K}+\mathcal{J}$, where 
$\mathcal{K}\rho\equiv -i(H_{\mathrm{eff}}\rho-\rho H_{\mathrm{eff}}^\dag)$ and $\mathcal{J}\rho\equiv \sum_{j=1}^L L_j\rho L_j^\dag$. 
The effective non-Hermitian Hamiltonian $H_{\mathrm{eff}}$ is given by $H_{\mathrm{eff}}\equiv H-\frac{i}{2}\sum_{j=1}^L L_j^\dag L_j$, 
and its explicit form is obtained by replacing $U$ in $H$ with $U-i\gamma$, thereby making the interaction strength complex-valued \cite{Durr09, GarciaRipoll09, Ashida16, Nakagawa18, Yamamoto19, Nakagawa19, Yoshida19}. 
Notably, the one-dimensional Hubbard model with a complex-valued interaction strength is still integrable \cite{Medvedyeva16, Nakagawa18, Ziolkowska19}. 
If the interaction strength becomes complex-valued, the SO(4) symmetry of the Hubbard Hamiltonian \cite{Yang89, YangZhang90, Essler91} remains intact. 
In particular, an eigenstate of the non-Hermitian Hubbard model can be labeled by the number of particles. 
Let $\ket{N,a}_R$ be a right eigenstate of $H_{\mathrm{eff}}$ with $N$ particles: $H_{\mathrm{eff}}\ket{N,a}_R=E_{N,a}\ket{N,a}_R$, where $a$ distinguishes the eigenstates having the same particle number. Then, one can diagonalize the superoperator $\mathcal{K}$ as $\mathcal{K}\varrho_{ab}^{(N,n)}=\lambda_{ab}^{(N,n)}\varrho_{ab}^{(N,n)}$, where $\lambda_{ab}^{(N,n)}\equiv -i(E_{N,a}-E_{N+n,b}^*)$ and $\varrho_{ab}^{(N,n)}\equiv\ket{N,a}_{R\ R}\bra{N+n,b}$. 
The superoperator $\mathcal{J}$ lowers the particle number but never increases it. Thus, in the representation with the basis $\{ \varrho_{ab}^{(N,n)}\}_{N,a,b}$, the Liouvillian $\mathcal{L}$ is a triangular matrix that can easily be diagonalized. 
This is a general property of Liouvillians of systems with loss \cite{Torres14}. 
Indeed, because the eigenvalues of a triangular matrix are given by its diagonal elements, the eigenvalues of the Liouvillian are given by $\lambda_{ab}^{(N,n)}$. The corresponding right eigenoperator is given by a linear combination of the basis as $C_{ab}^{(N,n)}\varrho_{ab}^{(N,n)}+\sum_{N'=0}^{N-2}\sum_{a',b'}D_{aba'b'}^{(N,N',n)}\varrho_{a'b'}^{(N',n)}$, where the coefficients $D_{aba'b'}^{(N,N',n)}$ are obtained from the matrix elements $\ _L\bra{N'-2,r}L_j\ket{N',r'}_R$ of the Lindblad operator $L_j$ with $\ket{N',r}_L$ being the left eigenstate dual to $\ket{N',r}_R$ \cite{Torres14, supple}. 
We thus conclude that if the non-Hermitian Hubbard Hamiltonian $H_{\mathrm{eff}}$ is integrable, the Liouvillian $\mathcal{L}$ is solvable. Note that this does not mean that the Liouvillian itself has an integrable structure such as the Yang-Baxter relation. Therefore, the mechanism of the solvability here is different from those of previous works on Yang-Baxter integrable Liouvillians \cite{Medvedyeva16, Shibata19_1, Shibata19_2, Ziolkowska19}.


\textit{Steady states}.--\ 
A steady state of the system is characterized by an eigenoperator of $\mathcal{L}$ with zero eigenvalue. 
If a state $\ket{\Psi}$ is a right eigenstate of $H_{\mathrm{eff}}$ with a real eigenvalue, one can show $L_j\ket{\Psi}=0$, and hence $\ket{\Psi}\bra{\Psi}$ is a steady state \cite{supple}. 
For example, the fermion vacuum $\ket{0}\bra{0}$ is trivially a steady state. 
Also, in the Hilbert subspace with no spin-down particles, all eigenstates of $H_{\mathrm{eff}}$ coincide with those in the noninteracting ($U=\gamma=0$) case and thus describe steady states. 
By letting the spin lowering operator act on the spin-polarized eigenstates, one can construct many steady states owing to the spin SU(2) symmetry of $H_{\mathrm{eff}}$, 
reflecting the fact that magnetization is conserved during the dynamics \cite{Buca12, Albert14}.  
Clearly, these steady states are ferromagnetic and far from the thermal equilibrium states of the one-dimensional Hubbard model. 
Physically, the steadiness of the ferromagnetic states can be understood from the Fermi statistics, because the spin wavefunction that is fully symmetric with respect to a particle exchange requires antisymmetry in the real-space wavefunction and forbids doubly occupied sites that cause a decay, as observed in Refs.~\cite{FossFeig12, Nakagawa19}. In general, a steady state realized after a time evolution becomes a statistical mixture of the above steady states that depends on the initial condition.


\textit{Bethe ansatz}.--\ 
We use the Bethe ansatz to obtain the eigenspectrum of the non-Hermitian Hubbard model $H_{\mathrm{eff}}$. The Bethe equations are \cite{LiebWu68, LiebWu03, 1dHubbard_book}
\begin{gather}
k_jL=\Phi+2\pi I_j-\sum_{\beta=1}^M\Theta\left(\frac{\sin k_j-\lambda_\beta}{u}\right),\label{eq_Bethe1}\\
-\sum_{j=1}^N \Theta\left(\frac{\sin k_j-\lambda_\alpha}{u}\right)=2\pi J_\alpha+\sum_{\beta=1}^M \Theta\left(\frac{\lambda_\alpha-\lambda_\beta}{2u}\right),\label{eq_Bethe2}
\end{gather}
where $N$ is the number of particles, $M$ is the number of down spins, $k_j\ (j=1,\cdots,N)$ is a quasimomentum, $\lambda_\alpha\ (\alpha=1,\cdots,M)$ is a spin rapidity, $u\equiv(U-i\gamma)/(4t)$ is a dimensionless complex interaction coefficient, and $\Theta(z)\equiv2\arctan z$. The quantum number $I_j$ takes an integer (half-integer) value for even (odd) $M$, and $J_\alpha$ takes an integer (half-integer) value for odd (even) $N-M$. Here we employ a twisted boundary condition $c_{L+1,\sigma}=e^{-i\Phi}c_{1,\sigma}$ for later convenience, but basically set $\Phi=0$ (i.e., the periodic boundary condition) unless otherwise specified.


\textit{Liouvillian gap}.--\ 
The late-stage dynamics of the system near a steady state is governed by long-lived eigenmodes whose eigenvalues are close to zero \cite{Cai13}. 
By construction of the steady states, the long-lived eigenmodes correspond to Bethe eigenstates in the $M=1$ case and their descendants derived from the spin SU(2) symmetry. 
They consist of ferromagnetic spin-wave-type excitations, and their dispersion relation is obtained by a standard calculation with the Bethe ansatz \cite{supple}. 
Taking consecutive charge quantum numbers $I_j=-(N+1)/2+j$, which express the simplest situation of charge excitations from the Fermi surface, we obtain an analytic expression for the dispersion relation of the spin excitations,
\begin{equation}
\Delta E\simeq -\frac{t}{\pi u}\left(Q_0-\frac{1}{2}\sin 2Q_0\right)\left(1-\cos\frac{\pi\Delta P}{Q_0}\right)
\label{eq_spinwave}
\end{equation}
for the momentum $\Delta P\simeq 0$, where $Q_0=\pi N/L$ is the Fermi momentum. 
Since the momentum is discretized in units of $2\pi/L$, the gapless quadratic dispersion around $\Delta P=0$ leads to the smallest imaginary part of the excitation energy $|\mathrm{Im}[\Delta E]|$ proportional to $1/L^2$. Thus, the Liouvillian gap, which is defined by the largest nonzero real part of eigenvalues of the Liouvillian, vanishes in the thermodynamic limit, implying a power-law time dependence in the decay dynamics \cite{Cai13}.

\begin{figure}
\includegraphics[width=8.5cm]{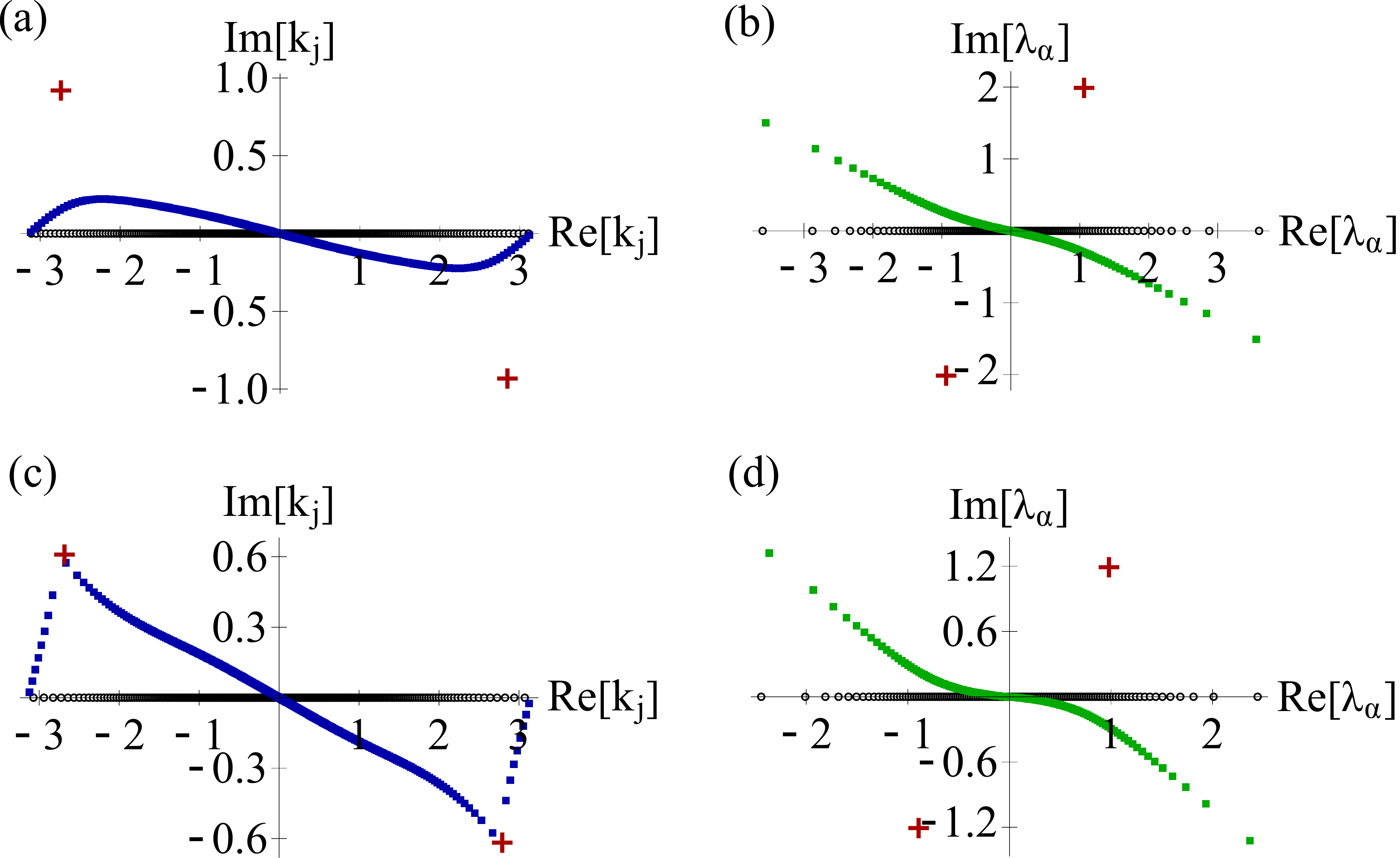}
\caption{Numerical solutions of the Bethe equations, Eqs.~\eqref{eq_Bethe1} and \eqref{eq_Bethe2}, for $L=N=2M=250$. (a),(c) Blue dots show quasimomenta $\{k_j\}$, and red crosses show the locations of poles at $k=\pm \pi-\arcsin(\pm iu)$.  (b),(d) Green dots show spin rapidities $\{\lambda_\alpha\}$, and red crosses show the locations of poles at $\lambda=\pm 2iu$. The interaction strength is set to $u=1-0.5i$ [(a),(b)] and $u=0.6-0.469i$ [(c),(d)]. Points on the real axis show the solutions for the case of $\gamma=0$ at the same $U$ for comparison.}
\label{fig_root}
\end{figure}

\textit{Hubbard gap, correlation length, and exceptional point}.--\ 
Next, we consider the half-filling case ($L=N=2M$) and focus on the solution that can be adiabatically connected to the ground state of the Hermitian Hubbard model in the limit of $\gamma\to 0$. Such a solution may not contribute to the late-stage behavior due to a short lifetime, but it can be used to study the early-time decay dynamics of a Mott insulator. 
We here assume that $U>0$ and $N\ (M)$ is even (odd), and set $I_j=-(N+1)/2+j$ and $J_\alpha=-(M+1)/2+\alpha$ as in the Hermitian case. 
In the thermodynamic limit, the Bethe equations, Eqs.~\eqref{eq_Bethe1} and \eqref{eq_Bethe2}, reduce to the integral equations for distribution functions $\rho(k)$ and $\sigma(\lambda)$ as
\begin{gather}
\rho(k)=\frac{1}{2\pi}+\cos k\int_\mathcal{S}d\lambda a_1(\sin k-\lambda)\sigma(\lambda),\label{eq_integral1}\\
\sigma(\lambda)=\int_\mathcal{C}dk a_1(\sin k-\lambda)\rho(k)-\int_\mathcal{S}d\lambda'a_2(\lambda-\lambda')\sigma(\lambda'),\label{eq_integral2}
\end{gather}
where $a_n(z)\equiv (1/\pi)[nu/(z^2+n^2u^2)]$, and $\mathcal{C}$ and $\mathcal{S}$ denote the trajectories of quasimomenta and spin rapidities, respectively \cite{1dHubbard_book}. 
Figure \ref{fig_root} (a),(b) show typical distributions of $\{ k_j\}_{j=1,\cdots,N}$ and $\{\lambda_\alpha\}_{\alpha=1,\cdots,M}$ that are obtained from the solution of the Bethe equations, Eqs.~\eqref{eq_Bethe1} and \eqref{eq_Bethe2}. 
The distributions indicate that if the trajectories $\mathcal{C}$ and $\mathcal{S}$ do not enclose a pole in the integrands of Eqs.~\eqref{eq_integral1} and \eqref{eq_integral2}, the trajectories can continuously be deformed to those of the $\gamma=0$ case, i.e., $\mathcal{C}=[-\pi,\pi]$ and $\mathcal{S}=(-\infty,\infty)$. Thus, we obtain the eigenvalue $E_0$ in the thermodynamic limit from analytic continuation of the solution in the $\gamma=0$ case \cite{LiebWu68} as
\begin{equation}
E_0/L=-2t\int_{-\infty}^\infty d\omega\frac{J_0(\omega)J_1(\omega)}{\omega(1+e^{2u|\omega|})},
\label{eq_GSenergy}
\end{equation}
where $J_n(x)$ is the $n$th Bessel function. Similarly, the Hubbard gap $\Delta_c$ \cite{LiebWu68, Woynarovich82_1} is given as
\begin{equation}
\Delta_c=4tu-4t\left[1-\int_{-\infty}^\infty d\omega\frac{J_1(\omega)}{\omega(1+e^{2u|\omega|})}\right].
\label{eq_Hubbardgap}
\end{equation}
Here $E_0$ and $\Delta_c$ take complex values in general. 
The lifetime of an eigenmode can be extracted from the imaginary part of the eigenvalue. The absolute value of $\mathrm{Im}[E_0]\leq 0$ first increases with increasing $\gamma$, takes the maximum at some point, and then decreases \cite{supple}. The decreasing behavior at large $\gamma$ is attributed to the continuous quantum Zeno effect \cite{Syassen08, Tomita17, Mark12, Barontini13, Yan13, Zhu14}, 
which prevents the creation of doubly occupied sites in eigenstates due to a large cost of the imaginary part of energy. 
By contrast, the absolute value of $\mathrm{Im}[\Delta_c]\leq 0$ monotonically increases with increasing $\gamma$ \cite{supple}, since the excitation corresponding to the Hubbard gap creates doubly occupied sites. 
As the Liouvillian eigenvalues appear as poles of a single-particle Green's function \cite{supple, Scarlatella19}, the dependence of the Hubbard gap on dissipation can be found from the linear response of the dynamics by, e.g., lattice modulation spectroscopy \cite{supple, Venuti16, Kollath06}.

\begin{figure}
\includegraphics[width=8.5cm]{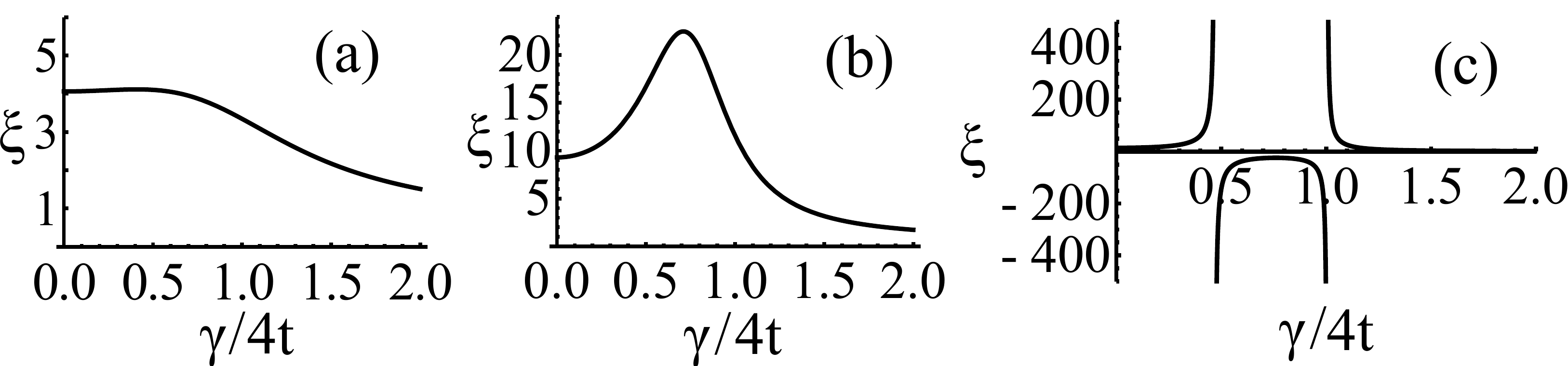}
\caption{Correlation length $\xi$ [Eq.~\eqref{eq_loc}] for (a) $U/4t=1$, (b) $U/4t=0.7$, and (c) $U/4t=0.6$.}
\label{fig_loc}
\end{figure}

To further elucidate the physics of the dissipative Mott insulator, we calculate the correlation length $\xi$ of the above eigenstate from the asymptotic behavior of the charge stiffness as $\Bigl|\frac{d^2E_0(\Phi)}{d\Phi^2}|_{\Phi=0}\Bigr|\sim\exp[-L/\xi]\ (L\to\infty)$ \cite{Stafford93}. 
The correlation length quantifies the dependence of the dynamics on the boundary condition and thus measures the spatial correlation in the eigenmode. 
We find that the correlation length is obtained from the analytic continuation of the result for the $\gamma=0$ case \cite{Stafford93}:
\begin{equation}
\frac{1}{\xi}=\mathrm{Re}\left[\frac{1}{u}\int_1^\infty dy\frac{\ln(y+\sqrt{y^2-1})}{\cosh(\pi y/2u)}\right].
\label{eq_loc}
\end{equation}
Figure \ref{fig_loc} (a)-(c) show the correlation length for different values of the repulsive interaction. For large $\gamma$, the correlation length decreases in all cases, indicating that particles are more localized due to dissipation. This behavior is consistent with the quantum Zeno effect \cite{Syassen08, Tomita17, Mark12, Barontini13, Yan13, Zhu14}. On the other hand, when $U$ is small, the correlation length grows at an intermediate dissipation strength [see Fig.~\ref{fig_loc} (b)], implying that dissipation facilitates the delocalization of particles. Surprisingly, the correlation length even diverges for small $U$ and takes negative values in between the divergence points [see Fig.~\ref{fig_loc} (c)]. 
When the correlation length diverges, the trajectory $\mathcal{C}$ crosses poles in the integrand of Eq.~\eqref{eq_integral2}, thereby preventing the trajectory from deforming to the real axis. 
This fact can be seen numerically (see red crosses on (off) the trajectory $\mathcal{C}$ ($\mathcal{S}$) in Fig.~\ref{fig_root} (c) [(d)]), and can also be shown analytically using the Bethe equations \cite{supple}. 
In fact, the solution with $\xi<0$ is not a solution of the Bethe equations, and the analytic continuation from the Hermitian case breaks down. 
Similar transitions of Bethe-ansatz solutions have been found in other non-Hermitian integrable models \cite{Albertini96, FukuiKawakami98, Nakagawa18}. 

The poles in the integrand in the first term on the right-hand side of Eq.~\eqref{eq_integral2} are given by $\sin k=\lambda\pm iu$. The same condition appears in the construction of the $k$-$\lambda$ string excitations in the Hubbard model \cite{1dHubbard_book, Takahashi72} in which a pair of quasimomenta $k^{(1)}, k^{(2)}$ form a string configuration around a center $\lambda$ as $\sin k^{(1)}=\lambda+iu$ and $\sin k^{(2)}=\lambda-iu$. Physically, such string excitations describe the creation of a doublon-holon pair \cite{1dHubbard_book}. The existence of the poles on trajectory $\mathcal{C}$ indicates that the solution in the thermodynamic limit becomes degenerate with a $k$-$\lambda$ string solution. In fact, the excitation energy of a $k$-$\lambda$ string is given by \cite{1dHubbard_book, Woynarovich82_1}
\begin{equation}
\varepsilon(k)=2tu+2t\cos k+2t\int_0^\infty d\omega\frac{J_1(\omega)\cos(\omega\sin k)e^{-u\omega}}{\omega\cosh u\omega},
\end{equation}
which vanishes at the poles $k=\pm\pi -\arcsin(\pm iu)$. Here not only the eigenvalues but also the eigenstates are the same. 
This means that the critical point at which the correlation length diverges is an exceptional point in the sense that the non-Hermitian Hamiltonian $H_{\mathrm{eff}}$ cannot be diagonalized \cite{Heiss12, AshidaGong20}. 
Importantly, we can show that the nondiagonalizability of $H_{\mathrm{eff}}$ leads to the nondiagonalizability of the Liouvillian $\mathcal{L}$ \cite{supple}. Thus, the exceptional point is the same for both the non-Hermitian Hamiltonian and the Liouvillian; however, this does not hold true for general Liouvillians \cite{Minganti19}. 
Since a nondiagonalizable Liouvillian leads to a singular time dependence of generalized eigenmodes \cite{Minganti19}, the exceptional point significantly alters the transient dynamics starting from half filling.

\begin{figure}
\includegraphics[width=8.5cm]{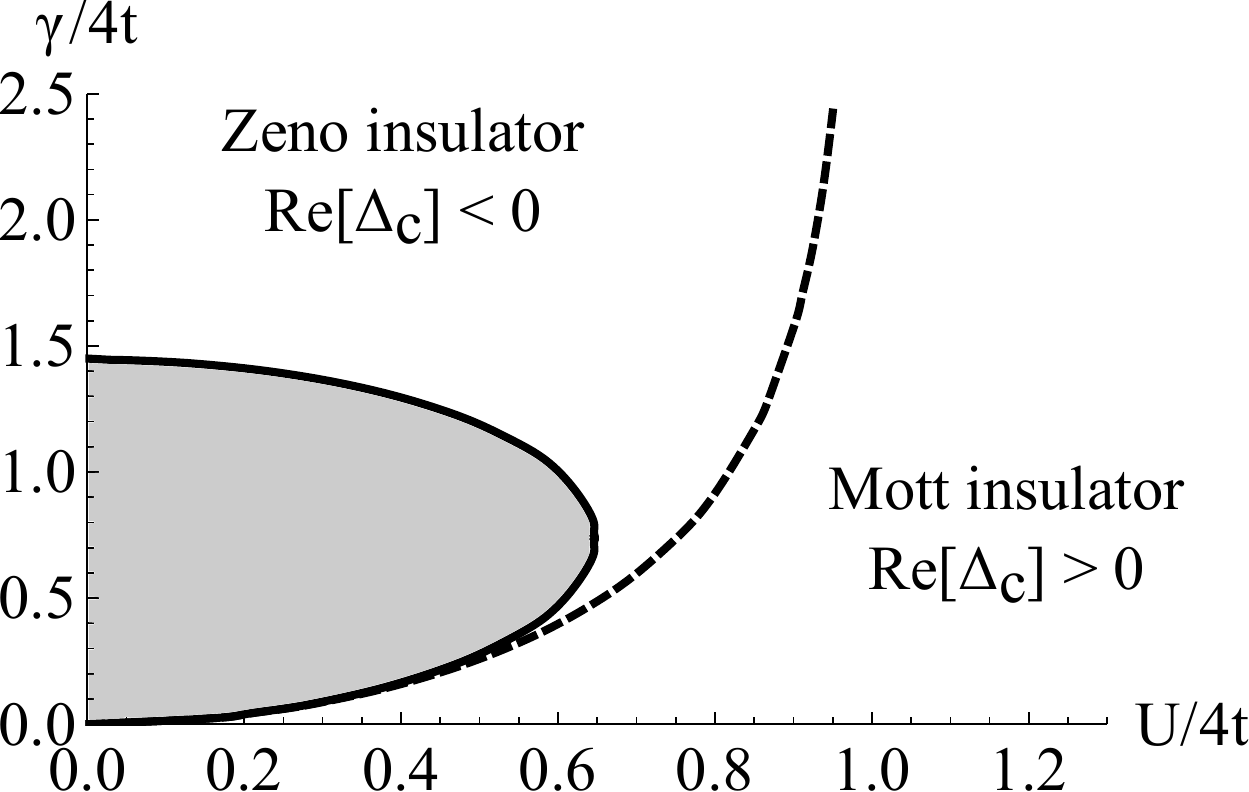}
\caption{``Phase diagram'' of the Liouvillian eigenmode that governs the transient dynamics at half filling. The solid curve indicates the location of the exceptional point at which the Liouvillian cannot be diagonalized. 
The shaded region cannot be analytically continued from the case with $\gamma=0$. 
The dashed curve shows where the real part of the Hubbard gap $\mathrm{Re}[\Delta_c]$ vanishes.}
\label{fig_phasediagram}
\end{figure}

The solid curve in Fig.~\ref{fig_phasediagram} shows the position of the exceptional point as a function of $U$ and $\gamma$. 
Outside the shaded region, the analytic continuation of the Bethe-ansatz solution from the $\gamma=0$ case remains valid. 
An increase of the correlation length in Fig.~\ref{fig_loc} (b) can be understood as a consequence of the proximity of the system to the exceptional point. 
For a large repulsive interaction $U>0$, a Mott insulator is formed as in the Hermitian Hubbard model and it has a finite lifetime due to nonzero $\gamma$. 
On the other hand, for small $U>0$ and large $\gamma$, particles are localized due to dissipation. Because the Hubbard gap becomes negative, $\mathrm{Re}[\Delta_c]<0$, in this region, the localization should be attributed to the quantum Zeno effect rather than the repulsive interaction, and therefore this localized state may be called a Zeno insulator. 
Interestingly, the phase diagram looks qualitatively similar to that obtained from a mean-field theory for a three-dimensional non-Hermitian attractive Hubbard model \cite{Yamamoto19} after changing the sign of $U$ via the Shiba transformation \cite{Shiba72}.

\textit{Dissipation-induced spin-charge separation}.--\ 
Finally, we address an interesting connection between strong correlations and dissipation. The Bethe equations, Eqs.~\eqref{eq_Bethe1} and \eqref{eq_Bethe2}, can be simplified when one takes the large-$|u|$ limit, in which one can expand the equations as (here we set $\Phi=0$)
\begin{gather}
k_jL=2\pi I_j +\mathcal{O}(1/u),\label{eq_Bethe3}\\
N\Theta\left(\frac{\lambda_\alpha}{u}\right)+\mathcal{O}(1/u^2)=2\pi J_\alpha+\sum_{\beta=1}^M\Theta\left(\frac{\lambda_\alpha-\lambda_\beta}{2u}\right).\label{eq_Bethe4}
\end{gather}
These equations indicate that the quasimomenta and spin rapidities are completely decoupled in the $|u|\to\infty$ limit \cite{Woynarovich82_1, OgataShiba90}. The quasimomenta in this limit are identical to those of free fermions, and Eq.~\eqref{eq_Bethe4} gives the same Bethe equation as that of the Heisenberg chain after rescaling $\Lambda_\alpha\equiv\lambda_\alpha/u$. This leads to a remarkable fact that the Bethe wavefunction is factorized into the charge part and the spin part \cite{OgataShiba90}. This argument is parallel to that for the spin-charge separation in the one-dimensional Hermitian Hubbard model. However, the crucial point here is that the spin-charge separation can occur due to large $\gamma$ even in the absence of the repulsive interaction $U$. Thus, in a Zeno insulator, the strong dissipation itself induces a strongly correlated state, and holes created by a loss behave as almost free fermions, whereas the spin excitations are described by a non-Hermitian Heisenberg chain with the exchange coupling $4t^2/(U-i\gamma)$ \cite{Nakagawa19}. As spin-charge separation in a Hermitian Hubbard chain has recently been observed in experiments with ultracold atoms \cite{Hilker17, Vijayan20}, the dissipation-induced spin-charge separation should be observed with current experimental techniques.


\textit{Conclusion}.--\ 
We have shown that the one-dimensional dissipative Hubbard model is exactly solvable. 
We have exploited the integrability of a non-Hermitian Hamiltonian to diagonalize a Liouvillian using the generic triangular structure of Liouvillians of systems with loss \cite{Torres14}. 
We have elucidated how strongly correlated states of the Hubbard model are fundamentally altered by dissipation, yet a number of important issues remain open. 
For example, the breakdown of the analytic continuation at half filling suggests that a novel state driven by an interplay between strong correlations and dissipation may be realized in the shaded region of Fig.~\ref{fig_phasediagram}. 
Since the standard solution for the Hermitian Hubbard model cannot be applied to that region, it is worthwhile to investigate the nature of Bethe-ansatz solutions with non-Hermitian interactions, as discussed in Refs.~\cite{Medvedyeva16, Ziolkowska19}. 
Finally, the solution of Liouvillians based on the non-Hermitian Bethe-ansatz method is not limited to the Hubbard model but applicable to other many-body integrable systems with appropriate Lindblad operators \cite{Torres14}. 
Examples include one-dimensional Bose \cite{LiebLiniger63_1, LiebLiniger63_2} and Fermi \cite{Yang67, Gaudin67} gases subject to particle losses \cite{Durr09}, quantum impurity models \cite{Andrei83, TsvelickWiegmann83} with dissipation at an impurity \cite{Nakagawa18}, and an XXZ spin chain \cite{YangYang66_1, YangYang66_2} with Lindblad operators that lower the magnetization \cite{Buca20}. 
We expect that the method proposed in this Letter can be exploited to uncover as yet unexplored exactly solvable models in open quantum many-body systems.

\begin{acknowledgments}
We are very grateful to Hosho Katsura for fruitful discussions. 
This work was supported by KAKENHI (Grant Nos.~JP18H01140, JP18H01145, JP19H01838, and JP20K14383) and a Grant-in-Aid for Scientific Research on Innovative Areas (KAKENHI Grant No.~JP15H05855) from the Japan Society for the Promotion of Science. 
\end{acknowledgments}

\textit{Note added}.--\ 
After the submission of this manuscript, a related work \cite{Buca20} appeared in which the Bethe-ansatz approach to triangular Liouvillians is studied for different models.

\bibliography{inelaHubbard_ref.bib}


\clearpage

\renewcommand{\thesection}{S\arabic{section}}
\renewcommand{\theequation}{S\arabic{equation}}
\setcounter{equation}{0}
\renewcommand{\thefigure}{S\arabic{figure}}
\setcounter{figure}{0}

\onecolumngrid
\appendix
\begin{center}
\large{Supplemental Material for}\\
\textbf{``Exact Liouvillian Spectrum of a One-Dimensional Dissipative Hubbard Model''}
\end{center}


\section{Eigensystem of the Liouvillian}
As mentioned in the main text, the eigenvalues of the Liouvillian $\mathcal{L}$ are given by $\lambda_{ab}^{(N,n)}=-i(E_{N,a}-E_{N+n,b}^*)$, where $E_{N,a}$ and $E_{N+n,b}$ are eigenvalues of the non-Hermitian Hubbard model $H_{\mathrm{eff}}$. The corresponding right eigenoperator $\rho_{ab}^{(N,n)}$ can be expanded in terms of the basis set $\{\varrho_{cd}^{(N',n)}=\ket{N',c}_{R\ R}\bra{N'+n,d}\}_{N',c,d}$ as 
\begin{equation}
\rho_{ab}^{(N,n)}=C_{ab}^{(N,n)}\varrho_{ab}^{(N,n)}+\sum_{N'=0}^{N-2}\sum_{a',b'}D_{aba'b'}^{(N,N',n)}\varrho_{a'b'}^{(N',n)}.
\label{eq_eigenvec}
\end{equation}
Note that the basis $\varrho_{cd}^{(N',n')}$ with $n'\neq n$ does not appear in the above expansion, since the U(1) charge symmetry of the Liouvillian enforces that the Liouvillian does not mix the sectors with different $n$ \cite{Buca12, Albert14}. 
Here, we follow Ref.~\cite{Torres14} to determine the coefficients $D_{aba'b'}^{(N,N',n)}$. We first expand the state in which a right eigenstate $\ket{N',a}_R$ is acted on by the Lindblad operator $L_j$ as
\begin{equation}
L_j\ket{N',a}_R=\sum_{r}v_{j,r}^{(N',a)}\ket{N'-2,r}_R,
\end{equation}
where $v_{j,r}^{(N',a)}=\ _L\bra{N'-2,r}L_j\ket{N',a}_R$ under the biorthonormal condition $\ _L\braket{N',a|N'',b}_R=\delta_{N',N''}\delta_{a,b}$. Then, we have
\begin{align}
\mathcal{J}\varrho_{ab}^{(N',n)}&=\sum_jL_j\varrho_{ab}^{(N',n)}L_j^\dag\notag\\
&=\sum_j\sum_{r,r'}v_{j,r}^{(N',a)}(v_{j,r'}^{(N'+n,b)})^*\ket{N'-2,r}_{R\ R}\bra{N'+n-2,r'}\notag\\
&=\sum_j\sum_{r,r'}v_{j,r}^{(N',a)}(v_{j,r'}^{(N'+n,b)})^*\varrho_{rr'}^{(N'-2,n)}.
\label{eq_Jmatele}
\end{align}
Substituting Eq.~\eqref{eq_eigenvec} into the eigenvalue equation $\mathcal{L}\rho_{ab}^{(N,n)}=\lambda_{ab}^{(N,n)}\rho_{ab}^{(N,n)}$, we obtain
\begin{align}
\mathcal{L}\rho_{ab}^{(N,n)}=&\lambda_{ab}^{(N,n)}C_{ab}^{(N,n)}\varrho_{ab}^{(N,n)}+\sum_j\sum_{r,r'}C_{ab}^{(N,n)}v_{j,r}^{(N,a)}(v_{j,r'}^{(N+n,b)})^*\varrho_{rr'}^{(N-2,n)}\notag\\
&+\sum_{N'=0}^{N-2}\sum_{a',b'}\lambda_{a'b'}^{(N',n)}D_{aba'b'}^{(N,N',n)}\varrho_{a'b'}^{(N',n)}+\sum_{N'=0}^{N-2}\sum_{a',b'}\sum_j\sum_{r,r'}D_{aba',b'}^{(N,N',n)}v_{j,r}^{(N',a')}(v_{j,r'}^{(N'+n,b')})^*\varrho_{rr'}^{(N'-2,n)},\label{eq_Lrho}\\
\lambda_{ab}^{(N,n)}\rho_{ab}^{(N,n)}=&\lambda_{ab}^{(N,n)}C_{ab}^{(N,n)}\varrho_{ab}^{(N,n)}+\sum_{N'=0}^{N-2}\sum_{a',b'}\lambda_{ab}^{(N,n)}D_{aba'b'}^{(N,N',n)}\varrho_{a'b'}^{(N',n)}.\label{eq_lambdarho}
\end{align}
Since the first terms on the right-hand side of Eqs.~\eqref{eq_Lrho} and \eqref{eq_lambdarho} are equal, we have
\begin{align}
\sum_j C_{ab}^{(N,n)}v_{j,a'}^{(N,a)}(v_{j,b'}^{(N+n,b)})^*+\lambda_{a'b'}^{(N-2,n)}D_{aba'b'}^{(N,N-2,n)}=\lambda_{ab}^{(N,n)}D_{aba'b'}^{(N,N-2,n)}
\end{align}
by comparing the coefficient of $\varrho_{a'b'}^{(N-2,n)}$. 
Thus, the coefficient $D_{aba'b'}^{(N,N-2,n)}$ is given by
\begin{equation}
D_{aba'b'}^{(N,N-2,n)}=\frac{1}{\lambda_{ab}^{(N,n)}-\lambda_{a'b'}^{(N-2,n)}}\left(\sum_j v_{j,a'}^{(N,a)}(v_{j,b'}^{(N+n,b)})^*\right)C_{ab}^{(N,n)}.
\label{eq_CN2}
\end{equation}
Similarly, by comparing the coefficient of $\varrho_{a'b'}^{(N-4,n)}$, we have
\begin{align}
\lambda_{a'b'}^{(N-4,n)}D_{aba'b'}^{(N,N-4,n)}+\sum_j\sum_{a'',b''}D_{aba''b''}^{(N,N-2,n)}v_{j,a'}^{(N-2,a'')}(v_{j,b'}^{(N+n-2,b'')})^*=\lambda_{ab}^{(N,n)}D_{aba'b'}^{(N,N-4,n)},
\end{align}
and hence
\begin{equation}
D_{aba'b'}^{(N,N-4,n)}=\frac{1}{\lambda_{ab}^{(N,n)}-\lambda_{a'b'}^{(N-4,n)}}\left(\sum_j\sum_{a'',b''}D_{aba''b''}^{(N,N-2,n)}v_{j,a'}^{(N-2,a'')}(v_{j,b'}^{(N+n-2,b'')})^*\right).
\end{equation}
Here, we have assumed $\lambda_{ab}^{(N,n)}-\lambda_{a'b'}^{(N',n)}\neq 0$ $(N'=0,2,\cdots,N-2)$, which is necessary for diagonalizability of the Liouvillian. 
Repeating the above procedures, all the coefficients $D_{aba'b'}^{(N,N',n)}\ (N'=0,\cdots,N-2)$ can be obtained recursively from $C_{ab}^{(N,n)}$. 

The left eigenoparator $\sigma_{ab}^{(N,n)}$ is obtained from the eigenvalue equation $\mathcal{L}^\dag\sigma_{ab}^{(N,n)}=(\lambda_{ab}^{(N,n)})^*\sigma_{ab}^{(N,n)}$, where the conjugate of the Liouvillian is defined by \cite{BreuerPetruccione}
\begin{equation}
\mathcal{L}^\dag\rho\equiv i(H_{\mathrm{eff}}^\dag\rho-\rho H_{\mathrm{eff}})+\sum_{j=1}^L L_j^\dag\rho L_j.
\end{equation}
The left eigenoperator can be expanded in terms of the basis set $\{\varsigma_{cd}^{(N',n)}=\ket{N',c}_{L\ L}\bra{N'+n,d}\}_{N',c,d}$ by using the left eigenstates as
\begin{equation}
\sigma_{ab}^{(N,n)}=\tilde{C}_{ab}^{(N,n)}\varsigma_{ab}^{(N,n)}+\sum_{N'=N+2}^{N_{\mathrm{max}}}\sum_{a',b'}\tilde{D}_{aba'b'}^{(N,N',n)}\varsigma_{a'b'}^{(N',n)},
\label{eq_eigenvec_left}
\end{equation}
where $N_{\mathrm{max}}=2L$ is the maximum value of the particle number of the system. One can determine the coefficients $\tilde{D}_{aba'b'}^{(N,N',n)}$ similarly as for the case of $D_{aba'b'}^{(N,N',n)}$ shown above. For example, 
\begin{align}
\tilde{D}_{aba'b'}^{(N,N+2,n)}=&\frac{1}{(\lambda_{ab}^{(N,n)})^*-(\lambda_{a'b'}^{(N+2,n)})^*}\left(\sum_j (v_{j,a}^{(N+2,a')})^*v_{j,b}^{(N+2+n,b')}\right)\tilde{C}_{ab}^{(N,n)},\\
\tilde{D}_{aba'b'}^{(N,N+4,n)}=&\frac{1}{(\lambda_{ab}^{(N,n)})^*-(\lambda_{a'b'}^{(N+4,n)})^*}\left(\sum_j\sum_{a'',b''}\tilde{D}_{aba''b''}^{(N,N+2,n)}(v_{j,a''}^{(N+4,a')})^*v_{j,b''}^{(N+4+n,b')}\right).
\end{align}
From Eqs.~\eqref{eq_eigenvec}, \eqref{eq_eigenvec_left}, the normalization condition for the eigenoperators reads
\begin{equation}
\mathrm{Tr}[(\sigma_{ab}^{(N,n)})^\dag\rho_{ab}^{(N,n)}]=(\tilde{C}_{ab}^{(N,n)})^*C_{ab}^{(N,n)}=1.
\end{equation}
The overall coefficients $C_{ab}^{(N,n)}$ and $\tilde{C}_{ab}^{(N,n)}$ in the expansions \eqref{eq_eigenvec} and \eqref{eq_eigenvec_left} can be fixed by using this normalization condition.

It follows from the above construction of the eigenoperators $\rho_{ab}^{(N,n)},\sigma_{ab}^{(N,n)}$ that if the non-Hermitian Hamiltonian $H_{\mathrm{eff}}$ lies at an exceptional point (i.e., it cannot be diagonalized), so does the Liouvillian $\mathcal{L}$. To see this, let us assume that the non-Hermitian Hamiltonian is parameterized as $H_{\mathrm{eff}}(g)$ and that it is at an exceptional point for $g=g_{\mathrm{EP}}$. The eigenvalue equation is given by $H_{\mathrm{eff}}(g)\ket{N,a,g}_R=E_{N,a}(g)\ket{N,a,g}_R$. At the exceptional point, at least two eigenstates and the corresponding eigenvalues are degenerate:
\begin{gather}
\lim_{g\to g_{\mathrm{EP}}}\left(E_{N,a_1}(g)-E_{N,a_2}(g)\right)=0,\\
\lim_{g\to g_{\mathrm{EP}}}\left(\ket{N,a_1,g}_R-\ket{N,a_2,g}_R\right)=0.
\end{gather}
Thus, we have
\begin{gather}
\lim_{g\to g_{\mathrm{EP}}}\left(\lambda_{a_1b}^{(N,n)}(g)-\lambda_{a_2b}^{(N,n)}(g)\right)=\lim_{g\to g_{\mathrm{EP}}}\left(\lambda_{ba_1}^{(N-n,n)}(g)-\lambda_{ba_2}^{(N-n,n)}(g)\right)=0,\\
\lim_{g\to g_{\mathrm{EP}}}\left(\varrho_{a_1b}^{(N,n)}(g)-\varrho_{a_2b}^{(N,n)}(g)\right)=
\lim_{g\to g_{\mathrm{EP}}}\left(\varrho_{ba_1}^{(N-n,n)}(g)-\varrho_{ba_2}^{(N-n,n)}(g)\right)=0,\\
\lim_{g\to g_{\mathrm{EP}}}\left(v_{j,r}^{(N,a_1)}(g)-v_{j,r}^{(N,a_2)}(g)\right)=0,
\end{gather}
where $\lambda_{ab}^{(N,n)}(g)=-i(E_{N,a}(g)-E_{N+n,b}^*(g)),\varrho_{ab}^{(N,n)}(g)=\ket{N,a,g}_{R\ R}\bra{N+n,b,g}$, and $v_{j,r}^{(N,a)}(g)=\ _L\bra{N-2,r,g}L_j\ket{N,a,g}_R$. 
Then, the above construction of the eigensystem of the Liouvillian shows that the Liouvillian eigenoperators corresponding to eigenvalues $\lambda_{a_1b}^{(N,n)}(g)$ and $\lambda_{a_2b}^{(N,n)}(g)$ are degenerate for $g=g_{\mathrm{EP}}$, indicating that the Liouvillian lies at an exceptional point. Note that the Liouvllian eigenoperators corresponding to eigenvalues $\lambda_{ba_1}^{(N-n,n)}(g)$ and $\lambda_{ba_2}^{(N-n,n)}(g)$ are also degenerate at $g=g_{\mathrm{EP}}$. Thus, at the exceptional point, the degeneracy of the eigenstates of the non-Hermitian Hamiltonian leads to a large number of degenerate eigenoperators of the Liouvillian.


\section{Proof of the statement on steady states}
We show that a steady state of the quantum master equation governed by the Liouvillian $\mathcal{L}$ is given by an eigenstate of the non-Hermitian Hamiltonian $H_{\mathrm{eff}}$ with a real eigenvalue. Let $\ket{\Psi}$ be a right eigenstate of the non-Hermitian Hamiltonian with a real eigenvalue: $H_{\mathrm{eff}}\ket{\Psi}=E\ket{\Psi}, E\in\mathbb{R}$. We assume that the eigenstate is normalized as $\braket{\Psi|\Psi}=1$. Then, we have
\begin{equation}
E=\bra{\Psi}H\ket{\Psi}-\frac{i}{2}\sum_{j=1}^L \bra{\Psi}L_j^\dag L_j\ket{\Psi}.
\end{equation}
Since $H$ and $L_j^\dag L_j$ are Hermitian operators, it is required from the real eigenvalue $E$ that
\begin{equation}
\sum_{j=1}^L\bra{\Psi}L_j^\dag L_j\ket{\Psi}=0.
\end{equation}
Since $\bra{\Psi}L_j^\dag L_j\ket{\Psi}\geq 0$, we obtain
\begin{equation}
\bra{\Psi}L_j^\dag L_j\ket{\Psi}=0
\end{equation}
$(j=1,\cdots,L)$, which is satisfied if and only if $L_j\ket{\Psi}=0$. Thus, we have
\begin{equation}
\mathcal{L}\ket{\Psi}\bra{\Psi}=0,
\end{equation}
which shows that $\ket{\Psi}\bra{\Psi}$ is a steady state of the quantum master equation.


\section{Derivation of the dispersion relation of spin-wave excitations}
\begin{figure}
\includegraphics[width=12cm]{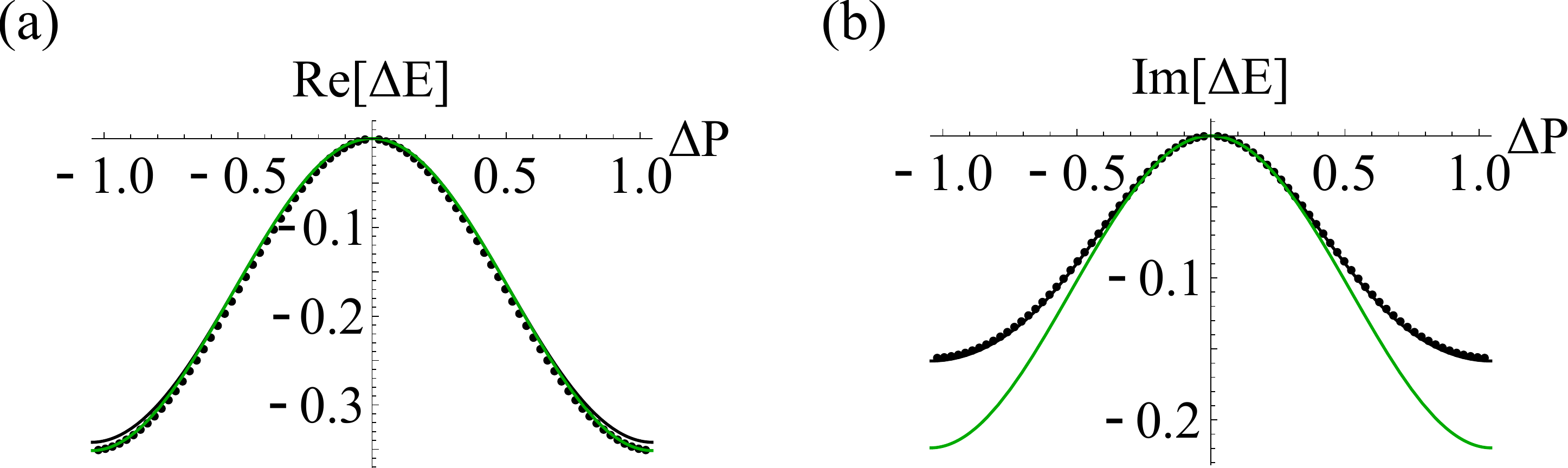}
\caption{Dispersion relation of the spin-wave-type excitations for $N/L=1/3$ and $u=0.8-0.5i$. (a) Real part of the excitation energy as a function of the momentum of an excitation. (b) Imaginary part of the excitation energy as a function of the momentum of the excitation. The dots are excitation energies calculated from numerical solutions of the Bethe equations \eqref{eq_Bethe1} and \eqref{eq_Bethe2} for $L=240$ and $N=80$. The black curves give dispersion relations obtained from Eqs.~\eqref{eq_excienergy} and \eqref{eq_excimom}. The green curves show approximate results [Eq.~\eqref{eq_approxdisp}] that become accurate for $\Delta P\simeq 0$.}
\label{fig_dispersion}
\end{figure}

Here, we derive the dispersion relation [Eq.~\eqref{eq_spinwave} in the main text] of spin-wave-type excitations which provide the long-lived eigenmodes during relaxation towards steady states. To this end, we consider the Bethe equations \eqref{eq_Bethe1} and \eqref{eq_Bethe2} in the main text with $M=1$. From Eq.~\eqref{eq_Bethe1}, we define the counting function $z_c(k)$ as
\begin{equation}
z_c(k)\equiv\frac{2\pi I_j}{L}=k+\frac{1}{L}\Theta\left(\frac{\sin k-\lambda_1}{u}\right).
\end{equation}
The distribution function $\rho(k)$ is then given by
\begin{align}
\rho(k)=&\frac{1}{2\pi}\frac{d z_c(k)}{dk}=\frac{1}{2\pi}+\frac{\cos k}{L}a_1(\sin k-\lambda_1)\notag\\
=&\rho_0+\frac{1}{L}\tilde{\rho}(k),
\end{align}
where $\rho_0\equiv 1/2\pi$ and $\tilde{\rho}(k)\equiv\cos k\cdot a_1(\sin k-\lambda_1)$. Here $\tilde{\rho}(k)$ gives a $1/L$ correction to the distribution function due to the excitation. 
The charge quantum numbers $\{ I_j\}_j$ specify the distribution of charge excitations. Since the long-lived eigenmodes are spin excitations, here we assume that the charge quantum numbers take consecutive values as $I_j=-(N+1)/2+j$, which express the simplest situation of charge excitations from the Fermi surface. 
Then, in the large-$L$ limit with a fixed density $N/L$, the quasimomenta are densely distributed over an interval $[-Q,Q]$, and $Q$ is determined from the particle density as
\begin{align}
\frac{N}{L}=&\int_{-Q}^{Q}dk\rho(k)\notag\\
=&\int_{-Q_0}^{Q_0}dk \rho_0+\frac{1}{L}\int_{-Q_0}^{Q_0}dk \tilde{\rho}(k)+2(Q-Q_0)\rho_0+\mathcal{O}(1/L^2),
\end{align}
where $Q_0=\pi N/L$. Thus, we have
\begin{equation}
Q-Q_0=-\frac{1}{L}\int_{-Q_0}^{Q_0}dk \frac{u\cos k}{(\sin k-\lambda_1)^2+u^2}+\mathcal{O}(1/L^2).
\end{equation}
The energy of an excitation is given by
\begin{align}
\Delta E=&-2tL\int_{-Q}^Q dk\rho(k)\cos k+2tL\int_{-Q_0}^{Q_0}dk\rho_0\cos k\notag\\
=&-2t\int_{-Q_0}^{Q_0}dk\tilde{\rho}(k)\cos k-2t\cdot 2(Q-Q_0)L\cdot\rho_0\cos Q_0\notag\\
=&-\frac{2t}{\pi}\int_{-Q_0}^{Q_0}dk\frac{u\cos k(\cos k-\cos Q_0)}{(\sin k-\lambda_1)^2+u^2},
\label{eq_excienergy}
\end{align}
and the momentum of the excitation is
\begin{align}
\Delta P=&\frac{2\pi}{L}\left(\sum_{j=1}^N I_j+J_1\right)-\frac{2\pi}{L}\sum_{j=1}^N\left(-\frac{N}{2}+j\right)\notag\\
=&-Q_0-\frac{1}{\pi}\int_{-Q_0}^{Q_0}dk\arctan\left(\frac{\sin k-\lambda_1}{u}\right).
\label{eq_excimom}
\end{align}
By eliminating $\lambda_1$ from Eqs.~\eqref{eq_excienergy} and \eqref{eq_excimom}, we obtain the dispersion relation. For $\Delta P\simeq 0$, the spin rapidity 
satisfies $|\lambda_1|\gg |\sin k|$, and hence $\Delta P\simeq -Q_0+\frac{2Q_0}{\pi}\arctan\frac{\lambda_1}{u}$. Thus, for $\Delta P\simeq 0$, we have
\begin{align}
\Delta E\simeq &-\frac{2t}{\pi}\int_{-Q_0}^{Q_0}dk(\cos^2k-\cos Q_0\cos k)\cdot\frac{u}{\lambda_1^2+u^2}\notag\\
\simeq&-\frac{t}{\pi u}\left(Q_0-\frac{1}{2}\sin 2Q_0\right)\left(1-\cos\frac{\pi\Delta P}{Q_0}\right),
\label{eq_approxdisp}
\end{align}
which is Eq.~\eqref{eq_spinwave} in the main text. In Fig.~\ref{fig_dispersion}, we plot the dispersion relation of the excitations for $-Q_0\leq \Delta P\leq Q_0$. 
When one takes the limit of $|u|\to\infty$, the same condition $|\lambda_1|\gg |\sin k|$ holds true. Thus, the dispersion relation \eqref{eq_approxdisp} becomes exact for all $\Delta P$ in the case of $|u|\to\infty$ (see also Ref.~\cite{Penc97}).

\section{Liouvillian spectrum as poles of single-particle Green's function}
The Hubbard gap in the main text is defined from eigenvalues of the non-Hermitian Hubbard model. To clarify its physical meaning, we consider a single-particle retarded Green's function $G^R_{j,j',\sigma}(\tau,\tau')$ defined by
\begin{equation}
G^R_{j,j',\sigma}(\tau,\tau')\equiv-i\theta(\tau-\tau')\langle\{ c_{j,\sigma}(\tau),c_{j',\sigma}^\dag(\tau')\}\rangle,
\end{equation}
where $c_{j,\sigma}(\tau)$ and $c_{j',\sigma}^\dag(\tau')$ are the Heisenberg representations of fermion annihilation and creation operators, the expectation value $\langle A\rangle\equiv\mathrm{Tr}[A\rho(0)]$ is taken over an initial state $\rho(0)$, $\{ A,B\}\equiv AB+BA$ is the anticommutator, and $\theta(\tau)$ is the Heaviside unit-step function. 
Suppose that the Liouvillian can be diagonalized. Then, the time evolution of the density matrix from time $\tau'$ to $\tau$ is expanded in terms of the eigenmodes as
\begin{align}
\rho(\tau)&=e^{\mathcal{L}(\tau-\tau')}\rho(\tau')\notag\\
&=\sum_N\sum_n\sum_{a,b}e^{\lambda_{ab}^{(N,n)}(\tau-\tau')}\frac{\mathrm{Tr}[(\sigma_{ab}^{(N,n)})^\dag \rho(\tau')]}{\mathrm{Tr}[(\sigma_{ab}^{(N,n)})^\dag \rho_{ab}^{(N,n)}]}\rho_{ab}^{(N,n)},
\end{align}
where $\rho_{ab}^{(N,n)}$ ($\sigma_{ab}^{(N,n)}$) is a right (left) eigenoperator of the Liouvillian with an eigenvalue $\lambda_{ab}^{(N,n)}$. Thus, the single-particle Green's function can be decomposed as
\begin{align}
G^R_{j,j',\sigma}(\tau,\tau')=&-i\theta(\tau-\tau')\sum_N\sum_n\sum_{a,b}e^{\lambda_{ab}^{(N,n)}(\tau-\tau')}\frac{\mathrm{Tr}[c_{j,\sigma}\rho_{ab}^{(N,n)}]\mathrm{Tr}[(\sigma_{ab}^{(N,n)})^\dag (c_{j',\sigma}^\dag\rho(\tau')+\rho(\tau')c_{j',\sigma}^\dag)]}{\mathrm{Tr}[(\sigma_{ab}^{(N,n)})^\dag \rho_{ab}^{(N,n)}]}.
\end{align}
After the Fourier transformation, we have
\begin{align}
G^R_{j,j',\sigma}(\tau';\omega)\equiv&\int_{-\infty}^\infty d(\tau-\tau')e^{(i\omega-\delta)(\tau-\tau')}G^R_{j,j',\sigma}(\tau,\tau')\notag\\
=&\sum_N\sum_n\sum_{a,b}\frac{w_{ab}^{(N,n)}(\tau')}{\omega-i\lambda_{ab}^{(N,n)}+i\delta},
\label{eq_Lehmann1}
\end{align}
which is a counterpart of the Lehmann representation of Green's function in open quantum systems \cite{Scarlatella19}. Here, 
\begin{equation}
w_{ab}^{(N,n)}(\tau')\equiv \frac{\mathrm{Tr}[c_{j,\sigma}\rho_{ab}^{(N,n)}]\mathrm{Tr}[(\sigma_{ab}^{(N,n)})^\dag (c_{j',\sigma}^\dag\rho(\tau')+\rho(\tau')c_{j',\sigma}^\dag)]}{\mathrm{Tr}[(\sigma_{ab}^{(N,n)})^\dag \rho_{ab}^{(N,n)}]},
\label{eq_w1}
\end{equation}
and $\delta=+0$ is an infinitesimal positive constant. Substituting the expansions \eqref{eq_eigenvec} and \eqref{eq_eigenvec_left} into Eq.~\eqref{eq_w1}, we obtain
\begin{align}
w_{ab}^{(N,n)}(\tau')=&\frac{1}{(\tilde{C}_{ab}^{(N,n)})^*C_{ab}^{(N,n)}}\left(C_{ab}^{(N,n)}{}_R\bra{N+n,b}c_{j,\sigma}\ket{N,a}_R+\sum_{N'=0}^{N-2}\sum_{a',b'}D_{aba'b'}^{(N,N',n)}{}_R\bra{N'+n,b'}c_{j,\sigma}\ket{N',a'}_R\right)\notag\\
&\times\Biggl( (\tilde{C}_{ab}^{(N,n)})^*{}_L\bra{N,a}(c_{j',\sigma}^\dag\rho(\tau')+\rho(\tau')c_{j',\sigma}^\dag)\ket{N+n,b}_L\notag\\
&+\sum_{N'=N+2}^{N_{\mathrm{max}}}\sum_{a',b'}(\tilde{D}_{aba'b'}^{(N,N',n)})^*{}_L\bra{N',a'}(c_{j',\sigma}^\dag\rho(\tau')+\rho(\tau')c_{j',\sigma}^\dag)\ket{N'+n,b'}_L\Biggr).
\end{align}
Clearly, $w_{ab}^{(N,n)}(\tau')=0$ for $n\neq -1$. Thus, we arrive at
\begin{align}
G^R_{j,j',\sigma}(\tau';\omega)=&\sum_N\sum_{a,b}\frac{1}{\omega-E_{N,a}+E_{N-1,b}^*+i\delta}\notag\\
&\times\left({}_R\bra{N-1,b}c_{j,\sigma}\ket{N,a}_R+\sum_{N'=0}^{N-2}\sum_{a',b'}(D_{aba'b'}^{(N,N',-1)}/C_{ab}^{(N,-1)}){}_R\bra{N'-1,b'}c_{j,\sigma}\ket{N',a'}_R\right)\notag\\
&\times\Biggl({}_L\bra{N,a}(c_{j',\sigma}^\dag\rho(\tau')+\rho(\tau')c_{j',\sigma}^\dag)\ket{N-1,b}_L\notag\\
&+\sum_{N'=N+2}^{N_{\mathrm{max}}}\sum_{a',b'}(\tilde{D}_{aba'b'}^{(N,N',-1)}/\tilde{C}_{ab}^{(N,-1)})^*{}_L\bra{N',a'}(c_{j',\sigma}^\dag\rho(\tau')+\rho(\tau')c_{j',\sigma}^\dag)\ket{N'-1,b'}_L\Biggr),
\label{eq_GF_result}
\end{align}
which shows that the Green's function has a pole at $\omega=E_{N,a}-E_{N-1,b}^*$. If one takes the limit of $\gamma\to 0$, the above expression reduces to the standard form for the closed system \cite{Scarlatella19} as
\begin{equation}
G^R_{j,j',\sigma}(\tau';\omega)=\sum_N\sum_{a,b}\frac{\bra{N-1,b}c_{j,\sigma}\ket{N,a}\bra{N,a}(c_{j',\sigma}^\dag\rho(\tau')+\rho(\tau')c_{j',\sigma}^\dag)\ket{N-1,b}}{\omega-E_{N,a}+E_{N-1,b}+i\delta},
\end{equation}
where $E_{N,a}$ and $\ket{N,a}$ are an energy eigenvalue and the corresponding eigenstate of the Hermitian Hubbard model. Thus, Eq.~\eqref{eq_GF_result} generalizes the standard expression in a closed system, and includes the effect of dissipation in eigenvalues of the non-Hermitian Hubbard model and in the weight $w_{ab}^{(N,-1)}(\tau')$. In particular, the real part of the Hubbard gap $\mathrm{Re}[\Delta_c]$ appears as a gap in the spectral function $-(1/\pi)\mathrm{Im}G^R_{j,j',\sigma}(\tau';\omega)$ as in the closed system. 
Similar Lehmann representations can generally be obtained for two-time correlation functions of physical quantities \cite{Scarlatella19}. This fact provides a way to experimentally measure the Hubbard gap of the non-Hermitian Hubbard model. For example, as done in cold-atom experiments in Refs.~\cite{Tomita17, Tomita18, Sponselee18}, one can prepare a Mott insulating state as an initial state, and let the dissipative dynamics start to evolve. Since a linear response of the Markovian dynamics to a probe field can be written with a two-time correlation function \cite{Venuti16}, the Hubbard gap may be observed in a spectroscopic signal by using, e.g., lattice modulation spectroscopy \cite{Kollath06}. The measurement of the Hubbard gap can also be used to distinguish the Mott and Zeno insulators as shown in Fig.~3 in the main text.


\section{Dependence of the Hubbard gap on dissipation}
\begin{figure}
\includegraphics[width=12cm]{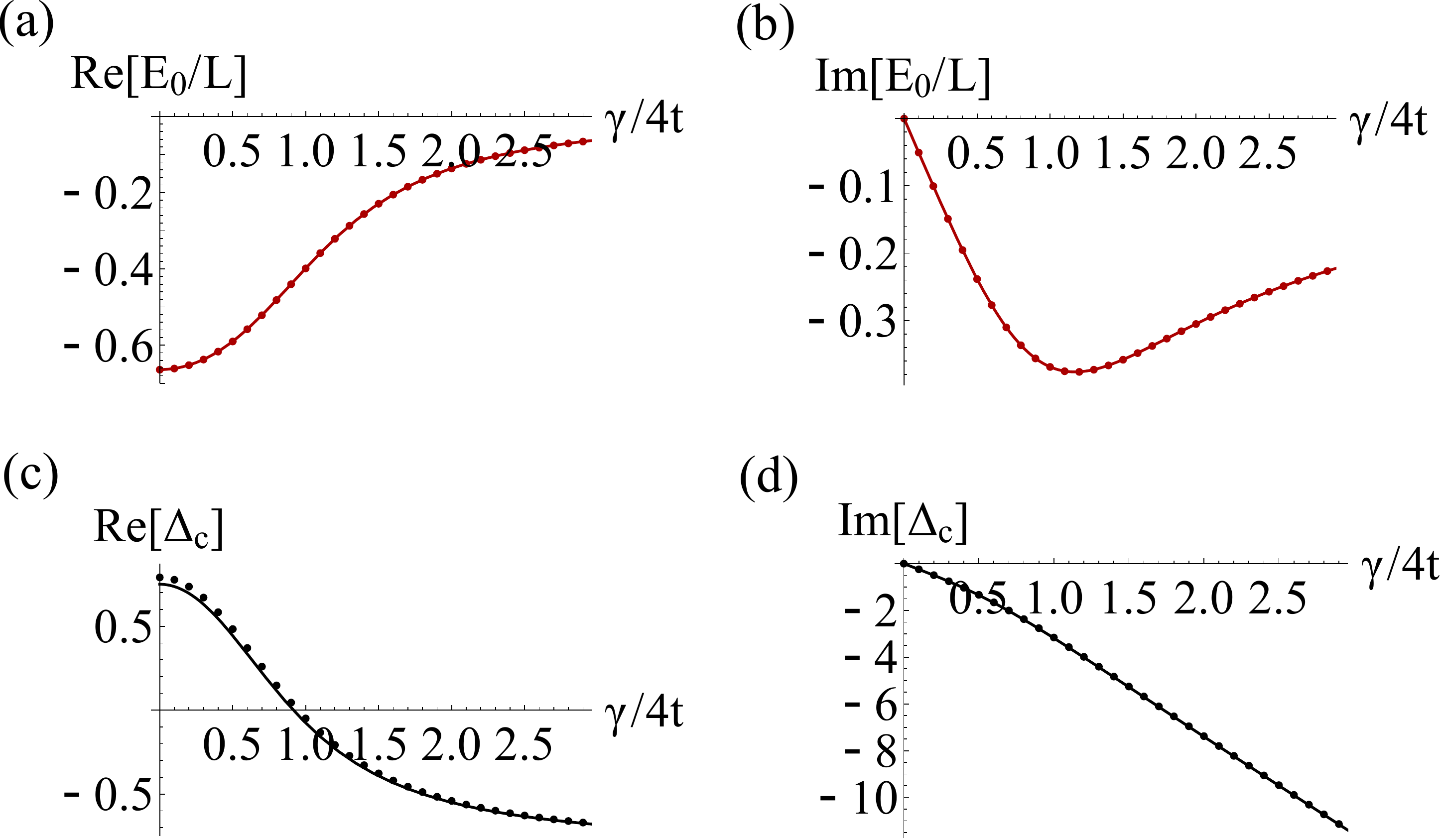}
\caption{(a) (b) Real [(a)] and imaginary [(b)] parts of the energy eigenvalue $E_0$ as a function of dissipation strength $\gamma$ for $U/4t=0.8$. Dots show numerical solutions of the Bethe equations \eqref{eq_Bethe1} and \eqref{eq_Bethe2} for $L=N=2M=50$. The solid curves are obtained from the analytic expression [Eq.~\eqref{eq_GSenergy}  in the main text] in the thermodynamic limit. (c) (d) Real [(c)] and imaginary [(d)] parts of the Hubbard gap $\Delta_c$ as a function of dissipation strength $\gamma$ for $U/4t=0.8$. Dots show numerical solutions of the Bethe equations \eqref{eq_Bethe1} and \eqref{eq_Bethe2} for $L=N=2M=50$. The solid curves are obtained from the analytic expression [Eq.~\eqref{eq_Hubbardgap} in the main text] in the thermodynamic limit.}
\label{fig_energy1}
\end{figure}

\begin{figure}
\includegraphics[width=12cm]{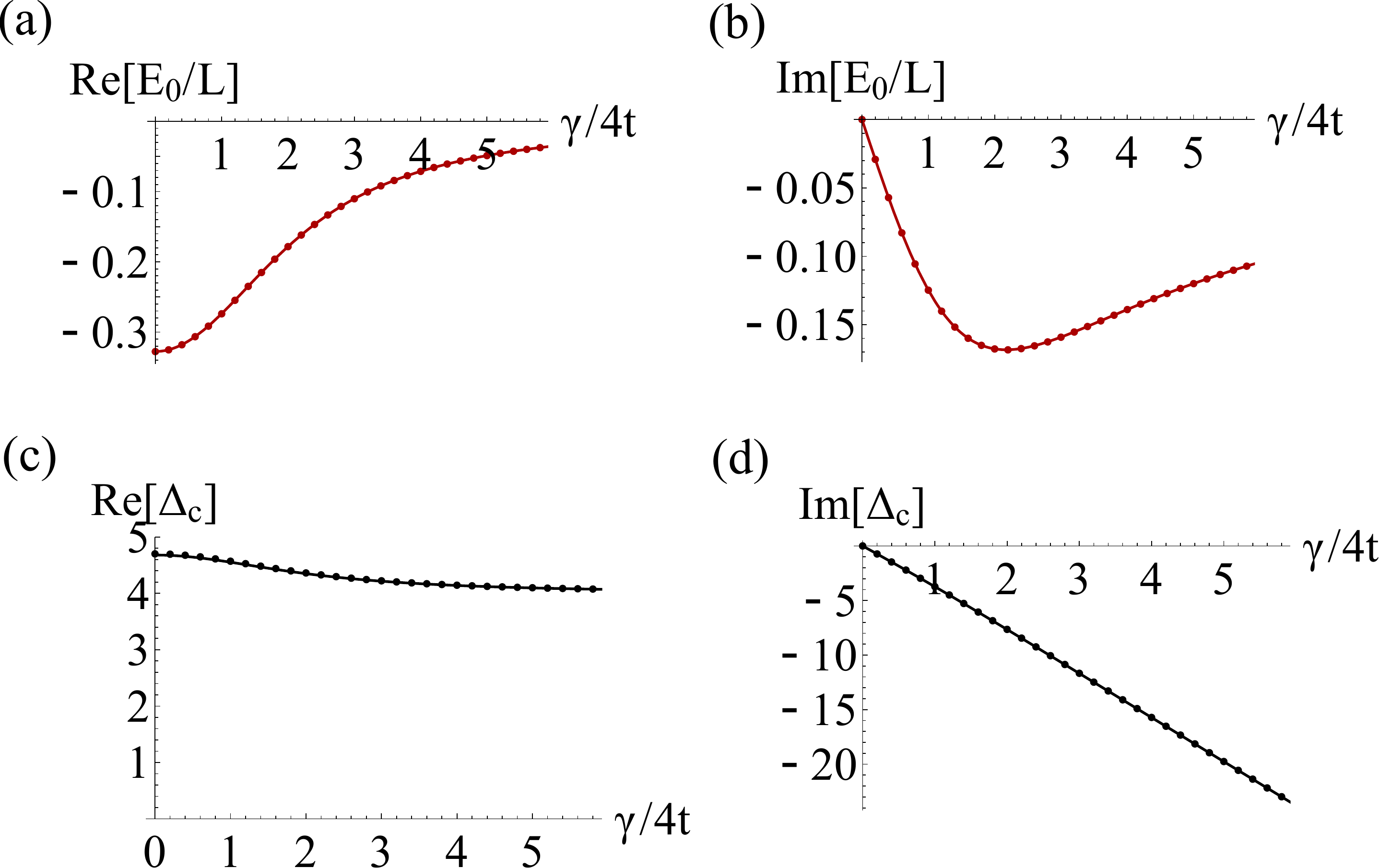}
\caption{(a) (b) Real [(a)] and imaginary [(b)] parts of the energy eigenvalue $E_0$ as a function of dissipation strength $\gamma$ for $U/4t=2$. Dots show numerical solutions of the Bethe equations \eqref{eq_Bethe1} and \eqref{eq_Bethe2} for $L=N=2M=50$. The solid curves are obtained from the analytic expression [Eq.~\eqref{eq_GSenergy}  in the main text] in the thermodynamic limit. (c) (d) Real [(c)] and imaginary [(d)] parts of the Hubbard gap $\Delta_c$ as a function of dissipation strength $\gamma$ for $U/4t=2$. Dots show numerical solutions of the Bethe equations \eqref{eq_Bethe1} and \eqref{eq_Bethe2} for $L=N=2M=50$. The solid curves are obtained from the analytic expression [Eq.~\eqref{eq_Hubbardgap} in the main text] in the thermodynamic limit.}
\label{fig_energy2}
\end{figure}

In Figs.~\ref{fig_energy1} and \ref{fig_energy2}, we show the dependence of the energy eigenvalue $E_0$ [Eq.~\eqref{eq_GSenergy} in the main text] and that of the Hubbard gap $\Delta_c$ [Eq.~\eqref{eq_Hubbardgap} in the main text] on dissipation for $U/4t=0.8$ and $U/4t=2$, respectively. The real part of the energy eigenvalue $\mathrm{Re}[E_0]$ monotonically increases with increasing the dissipation strength. The absolute value of the imaginary part of the energy eigenvalue $\mathrm{Im}[E_0]$ first increases with increasing $\gamma$, indicating that the dissipation causes a decay of the eigenmode. However, $|\mathrm{Im}[E_0]|$ reaches the maximum at an intermediate dissipation strength, and decreases for large $\gamma$. The decreasing behavior of $|\mathrm{Im}[E_0]|$ signals the onset of the quantum Zeno effect \cite{Syassen08, Tomita17, Mark12, Barontini13, Yan13, Zhu14}. While the qualitative behavior of the energy eigenvalue $E_0$ does not significantly depend on the magnitude of the repulsive interaction $U$, the Hubbard gap $\Delta_c$ shows a nontrivial dependence on $U$. For a weak repulsive interaction [see Fig.~\ref{fig_energy1} (c)], the real part of the Hubbard gap $\mathrm{Re}[\Delta_c]$ monotonically decreases with increasing the dissipation strength, and becomes negative when $\gamma$ exceeds a certain value. On the other hand, for a strong repulsive interaction [see Fig.~\ref{fig_energy2} (c)], the real part of the Hubbard gap remains positive for any $\gamma$. 
The qualitative difference of $\mathrm{Re}[\Delta_c]$ for small and large $U$ is attributed to the competition between the repulsive interaction and the quantum Zeno effect. For small $U$, particles are not well localized in the Mott insulator formed at $\gamma=0$, and therefore the Hubbard gap is significantly affected by localization due to the quantum Zeno effect. On the other hand, for large $U$, the Mott insulating state at $\gamma=0$ is not largely changed by dissipation, since particles are already well localized in the Mott insulator. Therefore, the real part of the Hubbard gap $\mathrm{Re}[\Delta_c]$ remains almost unchanged by increasing the dissipation strength. 
By contrast, the absolute value of the imaginary part of the Hubbard gap $\mathrm{Im}[\Delta_c]$ monotonically increases with increasing the dissipation strength [see Fig.~\ref{fig_energy1} (d) and Fig.~\ref{fig_energy2} (d)], because the excitation corresponding to the Hubbard gap creates doubly occupied sites and leads to $\mathrm{Im}[\Delta_c]\propto -\gamma$ for large $|u|$.

\section{Divergence of the correlation length at an exceptional point}
We follow Ref.~\cite{Stafford93} to calculate the correlation length $\xi$ as
\begin{equation}
\frac{1}{\xi}=\mathrm{Im}[z_c(k_*)],
\label{eq_xizc}
\end{equation}
where
\begin{equation}
z_c(k)=k+2\int_0^\infty d\omega\frac{J_0(\omega)\sin(\omega\sin k)}{\omega(1+e^{2u\omega})}
\end{equation}
is the counting function derived from the Bethe equations. Here $k_*=\pi-\arcsin(iu)$ denotes the stationary point of the counting function:
\begin{align}
\frac{dz_c}{dk}(k_*)=&1+2\cos k_*\int_0^\infty d\omega\frac{J_0(\omega)\cos(\omega\sin k_*)}{1+e^{2u\omega}}\notag\\
=&1-\sqrt{1+u^2}\int_0^\infty d\omega J_0(\omega)e^{-u\omega}\notag\\
=&0.
\end{align}
Note that $k_*$ also gives a pole of the integrand in the Bethe equation \eqref{eq_integral2} in the main text. Thus, if the pole $k_*$ is located on the trajectory $\mathcal{C}$ of quasimomenta, there exists a quasimomentum $k_j=k_*$ that satisfies
\begin{equation}
z_c(k_*)=\frac{2\pi I_j}{L}
\end{equation}
in the thermodynamic limit. Since the quantum number $I_j$ is real, the divergence of the correlation length $\xi=\infty$ occurs as can be seen from Eq.~\eqref{eq_xizc}.

\end{document}